\documentclass[12pt,preprint]{aastex}
\usepackage{graphicx}
\usepackage{epstopdf}

\newcommand{\eb}{\begin{equation}}
\newcommand{\ee}{\end{equation}}

\shorttitle{Tidal evolution of GJ 667c}
\shortauthors{Makarov et al.}
\begin{document}

\title{Dynamical evolution and spin-orbit resonances of potentially habitable exoplanets. The case of GJ 667C} 
\author{Valeri V. Makarov \& Ciprian Berghea}
\affil{United States Naval Observatory, 3450 Massachusetts Ave. NW, Washington DC, 20392-5420}
\email{vvm@usno.navy.mil}

\begin{abstract}
We investigate the dynamical evolution of the potentially habitable super-earth GJ 667Cc in the
multiple system of at least two exoplanets orbiting a nearby M dwarf, paying special attention to its
spin-orbital state. The published radial velocities for this star are re-analyzed and
evidence is found for additional periodic signals, which could be taken for two additional planets on
eccentric orbits. Such systems are not dynamically viable and break up quickly in numerical integrations.
The nature of the bogus signals in the available data remains unknown.
Limiting the scope to the two
originally detected planets, we assess the dynamical stability of the system and find
no evidence for bounded chaos in the orbital motion, unlike the previously investigated planetary
system of GJ 581.
The orbital eccentricity of the planets b and c is found to change cyclicly in the range 0.06 - 0.28
and 0.05 - 0.25, respectively, with a period of approximately 0.46 yr,
and a semimajor axis that little varies. Taking the eccentricity variation into account, numerical integrations are performed of 
the differential equations modeling the spin-orbit 
interaction of the planet GJ 667Cc with its host star, including fast oscillating components
of both the triaxial and tidal torques and assuming a terrestrial composition of its mantle.
Depending on the interior temperature of the planet, it is likely to be entrapped in the 3:2 (probability 0.51) or even
higher spin-orbit resonance. It is less likely to reach the 1:1 resonance (probability 0.24). Similar capture probabilities are obtained for the
inner planet GJ 667Cb. The estimated characteristic spin-down times are quite short for the two planets, i.e., within 1 Myr for
planet c and even shorter for planet b. Both planet arrived at their current and, most likely, ultimate spin-orbit
states a long time ago. The planets of GJ 667C are most similar to Mercury of all the Solar System bodies, as far as their tidal
properties are concerned. However, unlike Mercury, the rate of tidal dissipation of energy is formidably high in the planets of GJ 667,
estimated at $10^{23.7}$ and $10^{26.7}$ J yr$^{-1}$ for c and b, respectively. This raises a question of how such relatively
massive, close super-Earths could survive overheating and destruction.
\end{abstract}

\section{Introduction}
\label{firstpage}
\label{intro.sec}
According to a recent comprehensive study based on long-term spectroscopic observations with
HARPS, super-Earth exoplanets in the habitable zones of nearby M dwarfs are probably very
abundant, with an estimated rate of $\eta_{\rm Earth}=0.41^{+0.54}_{-0.13}$ per host
star \citep{bonf}. In exoplanet research, super-Earths usually designate planets
appreciably more massive than the Earth, but smaller than 10 Earth masses. From the same study,
the frequency of super-Earths with orbital periods between 10 and 100 days is $0.52^{+0.50}_{-0.16}$.
The current observational data thus do not rule out the possibility that there is a habitable
super-Earth in almost every M-type stellar system. This puts nearby M-stars into the focus
of exoplanet atmospheres and habitability research. In particular, detecting molecular tracers
of biological life no longer seems a merely speculative proposition.

The observations of planetary transits with {\it Kepler}, on the other hand, are better suited
to the detection of planets orbiting larger, solar-type stars. \citet{trau} estimates
that the frequency of terrestrial planets in the habitable zones of FGK stars is
$\eta_{\rm Earth}=0.34\pm{0.14}$. This estimate was obtained by extrapolation of the
observational statistics for planets with shorter periods ($<42$ days) collected from the
first 136 days of Kepler observations. Terrestrial planets may not be as common as ice giants,
but ubiquitous enough to investigate in earnest how extraterrestrial life can thrive in
other planetary systems.

Planet's rotation is one of the issues that have a bearing on habitability and atmosphere properties.
The major planets of the solar system, with the notable exception of Venus and Mercury, rotate
at significantly faster rates than their orbital motion. They are far enough from the Sun for the
tidal dissipation to be so slow that the characteristic times of spin-down are comparable to,
or longer, than the life time. The Earth's spin rate, for example, does slow down, but it will
take billions of years for the solar day to become appreciably longer. There is a common understanding
that the significance of tidal interactions is markedly larger in exoplanet systems. At the same time,
most publications on this topic were based on the two simplified, ``toy" models, one called the
constant-Q model and the other the constant time-lag model, which are either inaccurate for terrestrial
bodies or completely wrong \citep{ema}.

The stability of super-Earth's atmospheres around M dwarfs was investigated by \citet{heng}.
Their starting assumption, not justified by any dynamical consideration, was that such planets
are either synchronized (i.e., are captured into a 1:1 spin-orbit resonance) or pseudosynchronized.
The latter stable equilibrium state, when a planet rotates somewhat faster than the mean orbital motion,
is a direct prediction of the two above-mentioned simplified tidal models. Using a more realistic
tidal model developed in \citep{efr2,efrw}, \citet{mae} showed that a pseudosynchronous state is
inherently unstable for celestial bodies of terrestrial composition (the situation is less clear for
gaseous stars and liquid planets). Thus, a misassumption on the rotational state of a planet can nullify
the impact of an otherwise timely and significant theoretical study on planetary atmospheres.

The assumption that a close super-Earth should be synchronously rotating appears to be hardly assailable.
Do we not have a close analogy in the Moon, always facing the Earth with the same side? \citet{hut}
demonstrated that synchronously rotating components on a circular orbit is the only long-term
stable equilibrium in a two-body system with tides. This statement implies that given sufficiently long
time, all interacting systems should become synchronized and circularized. It appears entirely plausible
that the close super-Earths have had sufficient time to reach this terminal point of interaction.

This widely accepted point was critically reviewed in  \citep{mbe} where the possibly habitable super-Earth
GJ 581c was discussed. Compared to the rocky planets and moons of the Solar System, super-Earths orbiting
in the habitable zones of M stars are more massive, usually closer to their host stars, have mostly higher
orbital eccentricities, and are likely to be more axially symmetric. This combination of characteristics
makes them a very interesting class of objects for tidal dynamics. We would argue that the closest analogs
of super-Earths in our system are Mercury and Venus, rather than the Moon, Phobos or Titan. We found for
GJ 581d that the probabilities of capture into the supersynchronous resonances is so high with the observed eccentricity, that the planet
practically has no chance of reaching the 1:1 resonance, if its initial rotation was prograde. The most likely
state within a range of estimated parameters is a 2:1 spin-orbit resonance, when the planet makes exactly 2
complete sidereal rotations during one orbital revolution. 

In this paper, we offer a similar dynamical study of another exoplanet system, GJ 667C, which probably
includes at least two super-Earth planets. The composition of the system and the difficulties arising in the
interpretation of the spectroscopic data are discussed in \S\ref{orb.sec}. 
 
\section{The orbits of GJ 667C planets}
\label{orb.sec}
For our analysis of the orbital configuration of the GJ 667C system, we use the precision radial velocity
data published by \citet{delf}, collected with the HARPS instrument. The measurements span 2657 days and
have a median error of 1.3 m s$^{-1}$. The RV data obviously exhibit a systematic trend across the interval
of observation, rising by approximately 20 m s$^{-1}$. \citet{delf} note that the observed trend is consistent
with a gravitational acceleration exerted by the inner pair of stars (A and B) in this multiple star system.

Our planet detection algorithm is described in more detail in \citep{mbe}. Briefly, it is based on well-known and well-tested iterative 
periodogram analysis and brute-force grid search. The possible points of distinction with respect to other
algorithms used in exoplanet studies are:
1) we do not subtract the mean RV from the data but instead fit a constant term (and a trend when needed)
into the data in each iteration;
2) the original data are never altered in the process as each new signal is fitted along with all the previous
signals, gradually building up the model;
3) the nonlinear parameters of the Keplerian fit (eccentricity and phase) are optimized by a brute force grid search
minimizing the reduced $\chi^2$ statistics of the residuals. The confidence of each detection is computed through the
well-tested F-statistic, with the number of degrees of freedom decreasing by 7 with each planet signal added to the model.
Only detections with a confidence level above 0.99 are normally accepted. The algorithm was originally intended to be
a major part of a web-accessible multi-purpose tool for simulation, visualization and analysis of exoplanet systems,
constructed at NExScI (CalTech) in support of the {\it Space Interferometry Mission}. It was designed to be very fast and
to work on massive sets of data, optionally fitting combined RV and astrometric data. Error estimation was not a built-in
function of the algorithm, but rather was realized as a separate task based on the {\it Minimum Variance Bound} estimators.
For this reason, formal errors are not automatically provided for orbital fits, unless a dedicated statistical comparison
is in order.

\begin{figure}[htbp]
  \centering
  \includegraphics[angle=0,width=0.95\textwidth]{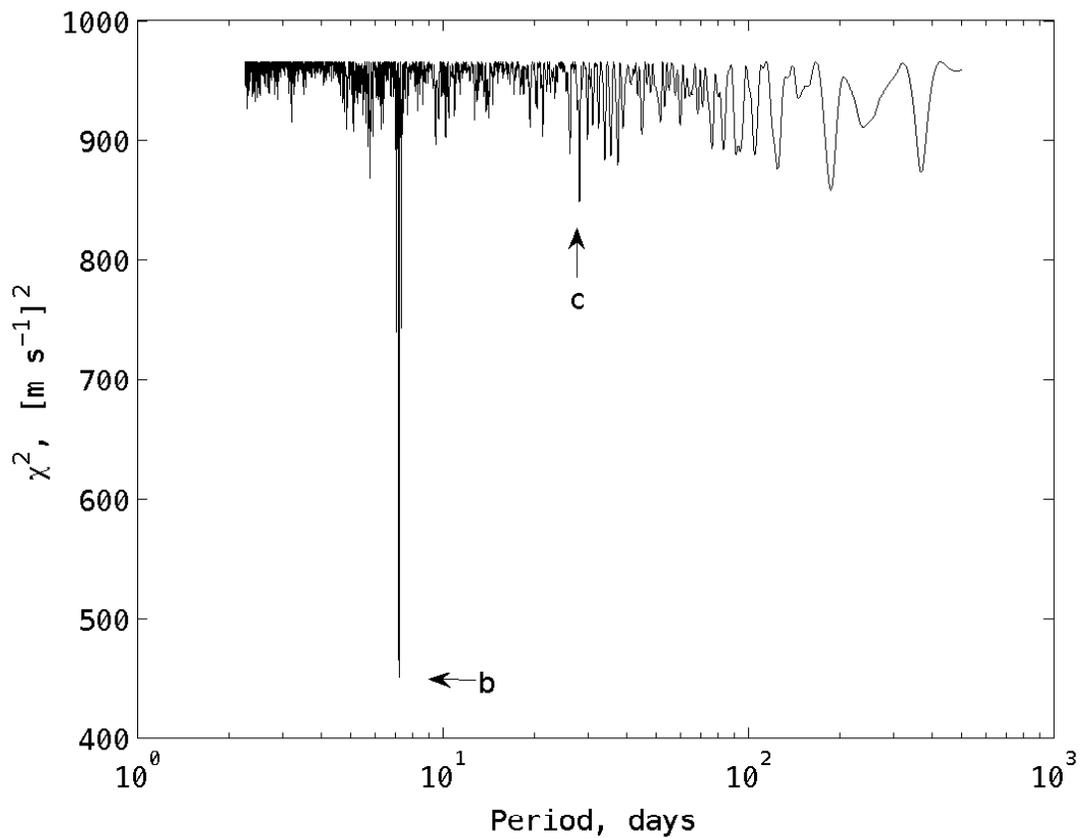}
\caption{The generalized $\chi^2$ periodogram of the GJ 667C system prior to orbital fitting
based on the HARPS data \citep{delf}.  The dips corresponding to the two confidently detected
planets are indicated with arrows and planet names. \label{per.fig}}
\end{figure}

Fig.~\ref{per.fig} shows the initial $\chi^2$ periodogram
of the HARPS data prior to any planet detection. The dips corresponding to the planets detected
from these data are marked with arrows and planet designations. The strongest signal is associated with planet
b, which is detected first. The second strongest signal corresponds to planet c. The
periodogram is riddled with a multitude of less significant dips of approximately equal strength, some of which are formally
accepted by the algorithm as bona fide planetary signals.
\begin{deluxetable}{rrrrrrrr}
\tablecaption{Orbital parameters of the two-planet GJ 667C system and additional signals
detected by our RVfit algorithm and by \citet{delf}. \label{ourfit.tab}}
\tablewidth{0pt}
\tablehead{
\multicolumn{1}{c}{Planet}  & \multicolumn{1}{c}{$P$} & \multicolumn{1}{c}{Mass} & \multicolumn{1}{c}{$a$} & \multicolumn{1}{c}{$e$} & \multicolumn{1}{c}{$\omega$} & \multicolumn{1}{c}{$M_0$}& \multicolumn{1}{c}{Note}\\
\multicolumn{1}{c}{}  & \multicolumn{1}{c}{d} & \multicolumn{1}{c}{$M_\earth$} & \multicolumn{1}{c}{AU} & \multicolumn{1}{c}{} & \multicolumn{1}{c}{$\degr$} & \multicolumn{1}{c}{$\degr$}& \multicolumn{1}{c}{}}
\startdata
\dotfill $b$ & $7.20$  & $5.80$ &   $0.050$ & 0.15   & 344 & 104 & 1\\
			& $7.199\pm0.001$ & 5.46 & 0.0504 & $0.09\pm 0.05$ & $-4\pm 33$ & & 2\\
\dotfill $c$ & $28.1$  & $4.28$ &   $0.125$ & 0.29   & 152 & 212 & 1\\
			& $28.13\pm0.03$ & 4.25 & 0.1251 & $0.34\pm 0.10$ & $166\pm 20$ & & 2\\
\dotfill $signal1$ & $91.5$ & $5.02$ & $0.275$ & 0.35 & 193 & 252 & 1\\
\dotfill $signal2$ & $53.3$  & $3.36$ & $0.192$ & 0.49 & 122 & 156 & 1\\
\dotfill $signal3$ & $35.3$  & $2.27$ & $0.145$ & 0.35 & 84 & 248 & 1\\
\enddata
\tablecomments{1: Our work; 2: \citet{delf}}
\end{deluxetable}

The orbital and physical parameters estimated from the fits of the five detected planets are given
in Table~\ref{ourfit.tab}. For the two safely detected planets b and c, our periods are in close agreement with the estimates by \citet{delf}
and \citet{angl}. A similarly very good agreement among the three studies is found for other parameters of b and c,
including the projected mass and semimajor axis, except for the eccentricity. As the eccentricity of planet b is almost
the same in this study and in \citet{angl} (0.15 and 0.17, respectively), it is appreciably smaller in \citet{delf} in
their 3-planet fit (Table 1, $0.09\pm 0.05$). On the contrary, planet's c eccentricity in \citet{angl} is restricted
to smaller values ($<0.27$) than those found in the other two studies (0.29 and 0.34). We conclude that there is a comfortable degree of
certainty about the properties of planets b and c, but the realistic uncertainty of eccentricities is probably up to 0.1.

This consistency breaks up completely when it comes to the interpretation of the remaining periodic signals in the HARPS data.
Different planet detection algorithms seem to pick up different sets of prominent features in the periodograms as
the most significant signals. \citet{angl} denote the residual feature as (d?) and find a period of 74.79 d for it,
forcing both the eccentricity and periastron longitude at zero. \citet{delf} offer a set of plausible two-planet solutions with
a linear trend, with the period of planet c taking values of 28, 90, 106, 124, 186, and 372 d, all yielding similar final
$\chi^2$ statistics on the residuals. Their baseline three-planet solution (with a trend) has a third planet d with a period
of 106.4 d and eccentricity 0.68. As we found out in our N-body simulations, such a system is not dynamically viable.
The authors also express strong doubts in the existence of a third planet, because they find a strong peak at a period of $\simeq105$
d in the activity diagnostics correlated with the rotation. A rotational period of 105 d is long, but not uncommon for 
inactive stars. Finally, our present analysis results in a set of features (labeled signal 1--3) with periods of 91.5, 53.3, and
35.3 d. Only the former periodicity seems to be in common with one of the tentative two-planet solutions of \citep{delf}. However,
we note that the 53.3 and 35.5 d periodicities are the second and third harmonics of the 106 d oscillation, which we do not explicitly
detect in our analysis. It would be interesting to investigate if a long-lived photospheric feature rotating with the star and
crossing the visible hemisphere with a period of 106 d could generate a whole set of harmonic signals in the RV periodogram.
It may be considered somewhat worrying that the period of planet c (28 d) is close to the fourth harmonic of this period.

The star GJ 667C is the tertiary component of a hierarchical multiple system, which is known to also include the K-dwarfs GJ 667A and B.
According to the Washington Double Star catalog (WDS), the separation between the A and C components was measured to 29.4 and 32.7
arcsec in 1875 and 2010, respectively. \citet{sod} obtained a much improved visual double star solution for components A and B, using
the Hipparcos observational data, and determined a semimajor axis of 1.81 arcsec, period 42.15 yr, and eccentricity 0.58. 
The D component listed in the WDS, on the other hand, is certainly optical, with a proper motion discordant with the fast-moving
GJ 667 system (William Hartkopf, priv. comm.). Thus, no companions more distant than C are known in this system. The considerable
eccentricity of the AB pair could be pumped by the C component via the Kozai-cycle mechanism of orbital interaction. On the other hand,
the observed trend in the RV of the C component can be caused by its orbital acceleration around the AB pair. 

\section{Long-term evolution of the orbits}
\label{cha.sec}

\begin{figure}[htbp]
\plotone{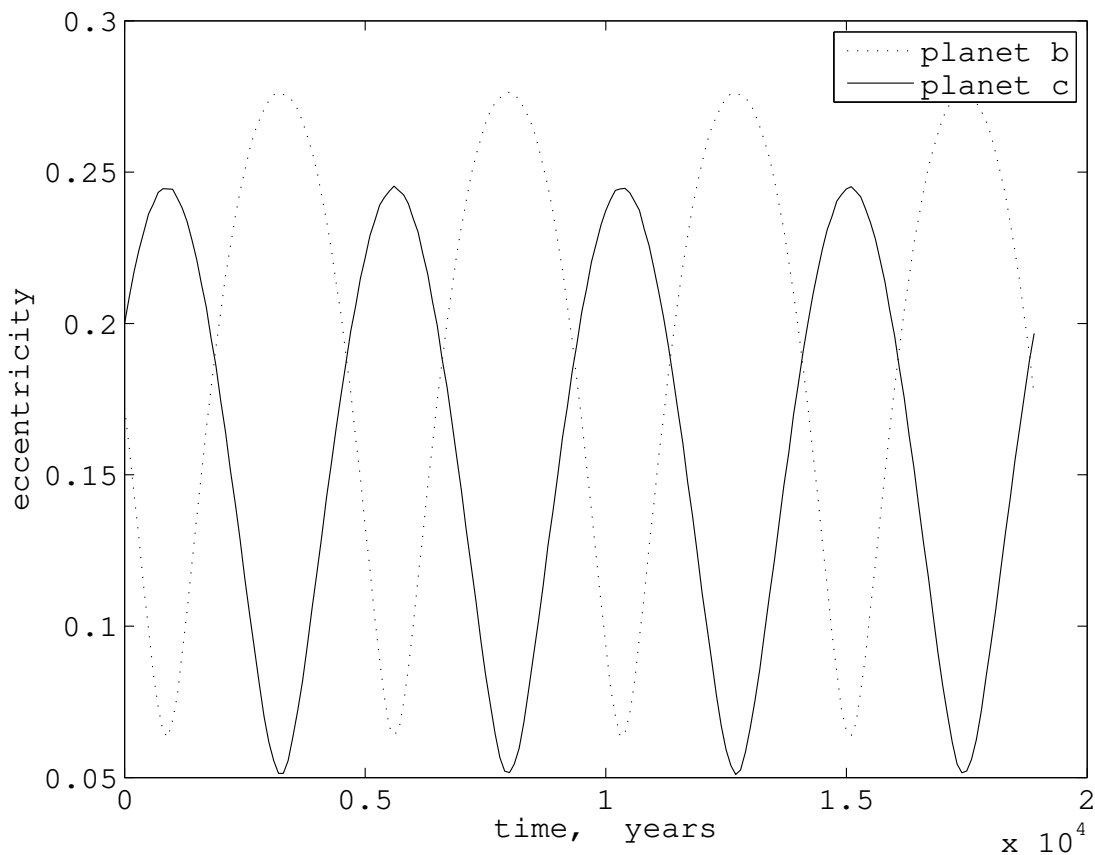} \epsscale{0.75}
\caption{Simulated eccentricity variations of planets GJ 667Cc and GJ 667Cd.   \label{ecc.fig}}
\end{figure}

Using the values listed in Table \ref{para.tab} as initial parameters and assuming the orbits to be coplanar we
performed multiple simulations of the dynamical evolution of a 2-body GJ 667C system. The parameters given
in this Table are consistent with the data, considered to be the best Keplerian solution in \citep{angl},
but neglecting the third, tentative planet d. The only change made is the eccentricity of planet c, which was set to
0.20. We chose to adopt the results from {\it ibid}, because our own Keplerian fit (Table \ref{ourfit.tab}) did not
seem to provide more reliable or accurate information about the reliably detected planets b and c. In order to
investigate the dynamical stability of such a three-body system, we employed
the symplectic integrator HNBody, version 1.0.7, \citep{rauch} with the hybrid symplectic
option.
The time step was set at  0.036 hours and the orbits of the planets were integrated for 5 million years
or longer. The mass of the star was assumed to be $0.33\,M_{\sun}$.

We find that the system of two planets (b and c) is perfectly stable. As found already by \citep{angl}, the
system is stabilized by a close 4:1 mean-motion resonance (MMR) with the arguments of periastron librating around
$180\degr$. The semimajor axes vary little.
The eccentricities show significant periodic variations with opposite phases, so that when planet b reaches
the largest eccentricity, planets c assumes the smallest eccentricity, and vice versa, see Fig. \ref{ecc.fig}.
These variations for both planets for the first 20000 years of integration are shown in Fig. \ref{ecc.fig}. 
The range of variations of eccentricity for planet b is between 0.06 and 0.27, and for planet c,
between 0.05 and 0.25. The period of libration in eccentricity is approximately 0.47 yr.

\begin{deluxetable}{rrrrrrr}
\tablecaption{Orbital parameters of the two-planet GJ 667C system used for orbital integration in this paper. \label{angl.tab}}
\tablewidth{0pt}
\tablehead{
\multicolumn{1}{c}{Planet}  & \multicolumn{1}{c}{$P$} & \multicolumn{1}{c}{Mass} & \multicolumn{1}{c}{$a$} & \multicolumn{1}{c}{$e$} & \multicolumn{1}{c}{$\omega$} & \multicolumn{1}{c}{$M_0$}\\
\multicolumn{1}{c}{}  & \multicolumn{1}{c}{d} & \multicolumn{1}{c}{$M_\earth$} & \multicolumn{1}{c}{AU} & \multicolumn{1}{c}{} & \multicolumn{1}{c}{$\degr$} & \multicolumn{1}{c}{$\degr$}}
\startdata
\dotfill $b$ & $7.20$  & $5.68$ &   $0.049$ & 0.17   & 344 & 107\\
\dotfill $c$ & $28.16$  & $4.54$ &   $0.123$ & 0.20   & 238 & 144\\
\enddata
\end{deluxetable}
 
\begin{figure}[htbp]
  \centering
  \plottwo{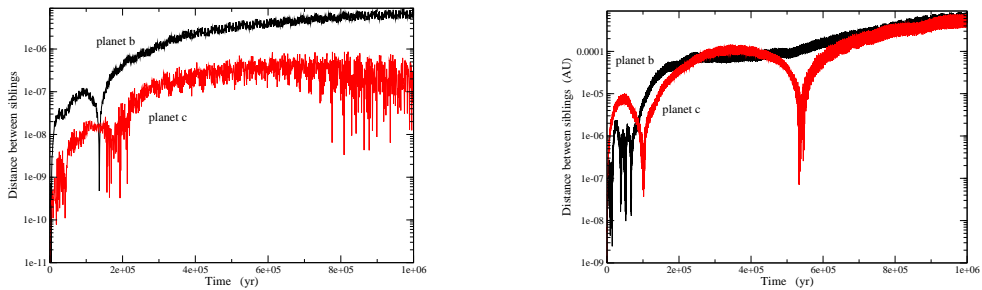}{twostars.eps}
\caption{Absolute distance between two sibling integrations of the orbits of planets GJ 667Cc and GJ 667Cd.  Left: three
bodies simulated in the system, i.e., the host star GJ 667C and the planets. Right: four bodies simulated, a single
primary body replacing the close pair of stars GJ 667 A and B, the host star GJ 667C, and the two planets.\label{chaos.fig}}
\end{figure}

We also numerically probed for evidence of chaos in the orbital motion of GJ 667C, employing the sibling simulation technique proposed by
\citet{hay1,hay2} to investigate the behavior of the outer Solar system. The orbits of the two planets were integrated with slightly different 
initial conditions for up to 5 million years. 
Two sibling trajectories are generated by perturbing the initial semi-major axis of the planet GJ 667Cb
by a factor of $10^{-14}$.
The distance between the unperturbed planets and their siblings is then computed as a function of time.
A chaotic motion manifests itself as an exponential divergence between the trajectories.
We did not find any signs of chaos in the motion of this system, the difference between the sibling
trajectories being polynomial over the entire span of integration (Fig. \ref{chaos.fig}, left). This marks 
a significant difference with
the previously studied case of GJ 581, where a rapid onset of chaos with Lyapunov times of tens to a hundred years
was unambiguously detected. The important difference between the system of GJ 581 and the system of GJ 667C
is that the former includes more than two planets, which may significantly interact with one another, whereas the
latter, as integrated in this paper, includes only two planets in 4:1 mean motion resonance. This result should thus be taken with
a caveat, because we do not know if no other planets orbit GJ 667C. Recalling that the host star is a tertiary in a hierarchical
triple system, we investigated if the presence of other distant and massive bodies could invoke dynamical chaos in the resonant
planetary system. The pair of stars A and B was replaced in these simulation by a single point mass of double the estimated mass
of the C component. This primary mass was placed at 297 AU from C with some randomly selected initial mean anomaly, an
inclination of $30\degr$ and zero eccentricity. Still no chaos was detected with the sibling method 
(Fig. \ref{chaos.fig}, right), at least within the
first $8\times 10^5$ yr. To rule out possible technical errors, this result was verified with an independent integration technique, for
which the well-tested Mercury code was selected \citep{cha}. The conclusion drawn from these numerical simulations is that GJ 667C with 
its two confidently detected planets represents a remarkably
stable system in the 4:1 MMR with a rapid periodical exchange of angular momentum between the two planets.

\vspace{1 cm}

\begin{deluxetable}{lrr}
\tablecaption{Default parameters GJ 667C planets b and c.\label{para.tab}}
\tablewidth{0pt}
\tablehead{
\multicolumn{1}{c}{Parameter}  &
\multicolumn{1}{c}{b} & \multicolumn{1}{c}{c}\\
}
\startdata
$\xi$ & \dotfill $0.4$ & 0.4\\[1ex]
$R$ & \dotfill $1.79 R_{\rm Earth}$ & $1.65 R_{\rm Earth}$\\
$M_2$ & \dotfill $5.7 M_{\rm Earth}$ & $4.5 M_{\rm Earth}$\\
$a$ & \dotfill  $0.049$ AU & $0.12$ AU \\
$e$ & \dotfill 0.27 & 0.25\\
$(B-A)/C$ & \dotfill $5\cdot 10^{-5}$ & $5\cdot 10^{-5}$\\
$P_{\rm orb}$ & \dotfill $7.2$ d & 28.1 d\\
$\tau_M$ & \dotfill 50 yr & 50 yr\\
$\mu$ & \dotfill $0.8\cdot10^{11}$ kg m$^{-1}$ s$^{-2}$ & $0.8\cdot10^{11}$ kg m$^{-1}$ s$^{-2}$\\
$\alpha$ & \dotfill $0.2$ & 0.2\\
\enddata
\end{deluxetable}

\section{The likely spin-orbit state of GJ 667Cc}
\label{c.sec}
\begin{figure}[htbp]
  \centering
  \plottwo{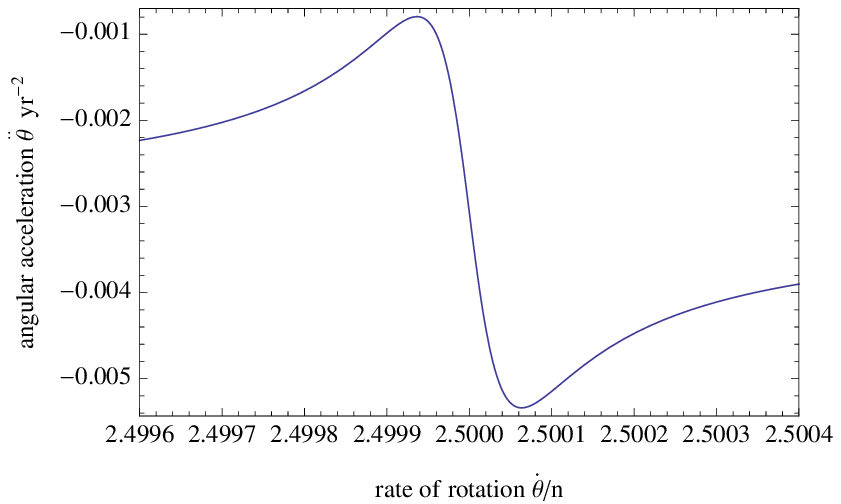}{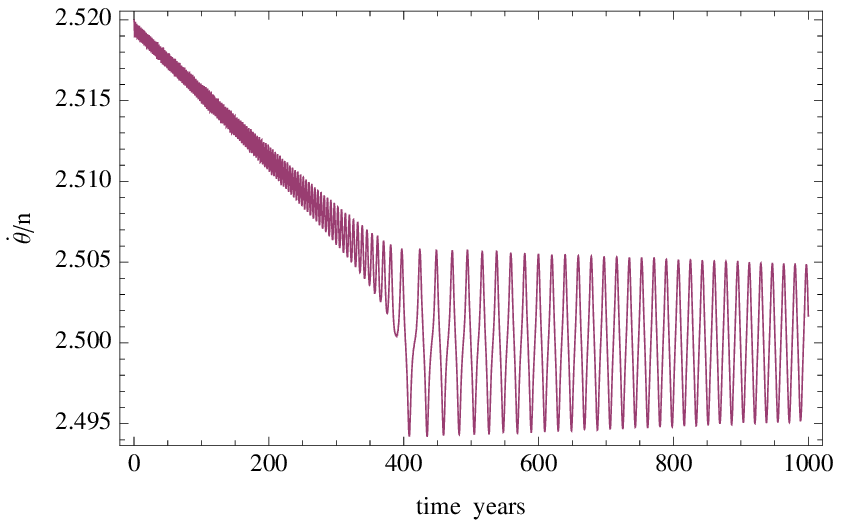}
\caption{Left: Secular tidal acceleration of GJ 667Cc versus normalized spin rate in the vicinity of the 5:2
spin-orbit resonance. Right: A simulated capture of GJ 667Cc into the 5:2 resonance.\label{kink.fig}}
\end{figure}

The polar torque acting on a rotating planet is the sum of the gravitational torque, caused by
the triaxial permanent shape and the corresponding quadrupole inertial momentum, and the tidal torque, caused by the dynamic deformation of
its body. The former torque is often considered to be strictly periodical, resulting in the net zero
acceleration of rotation. An important violation of this rule is discussed in \citep{makti}, namely, that when the planet
is locked in a spin-orbit resonance, a compensating nonzero secular torque emerges. The suitable equation for the
oscillating triaxial torque can be found in, e.g., \citep{danb,gold,gol68}. For the tidal component of torque,
we are using an advanced model based on the development of Kaula and Darwin's harmonic decomposition in \citep{efr2,efrw,efr1,efrb}, 
combined with a rheological model approximately known for the Earth. The rheological model derives from two principles, which
we believe are valid for any silicate or icy bodies: 1) in the low-frequency limit, the mantle's behavior should be close to that
of the Maxwell body because of the dominant mechanism of dissipation, namely, the lattice diffusion; 2) at frequencies higher than a
certain threshold (which is $\simeq 1$ yr$^{-1}$ for Earth), the mantle behaves as the Andrade body due to the pinning-unpinning of
dislocations. The threshold may move depending on the temperature of the mantle, which is unknown for exoplanets. Other rheological
parameters may vary too, depending on the average temperature and chemical constituency. However, the qualitative shape of the
frequency-dependence of tidal response shown in Fig. \ref{kink.fig} (left) should be universal for rocky or icy planets.
Possible deviations from this model may be caused by extensive internal or surface oceans, a subject outside the scope of this paper.

As has been emphasized in the previous publications about this
theory, profound implications for planets and moons of terrestrial composition call for a significant review of currently
accepted views and assumptions \citep{mae,ema}. In particular, capture into higher-order, supersynchronous resonances is much
easier than in the previously widely exploited models of linear or constant (frequency-independent) torque. This is due to the
kink-shape of the secular tidal torque component, which acts as an efficient trap abruptly arresting the gradual spin-down. Fig. \ref{kink.fig}(left)
shows the dependence of the secular tidal acceleration for GJ 667Cc with normalized spin rate $\dot\theta/n$ in a narrow vicinity of the
5:2 resonance. The default parameters used in this calculation, and in most of our other simulations, are given in Table \ref{para.tab}.

As explained in more detail in \citep{makti, noel}, the structure of secular torque in the vicinity of a spin-orbit resonance can be considered
to consist of two functionally different components: the resonant kink, which is perfectly symmetrical around the point of resonance, and
a nearly constant bias. In fact, the bias is nothing other than the sum of the distant parts of all other resonant kinks, separated by $1\,n$
in $\dot\theta$. For low and moderate eccentricities, the amplitude of the kinks is a rapidly decreasing function of the resonance order $q$,
the synchronous kink ($lmpq=2200$) being by far the dominating one. Therefore, the bias at the 1:1 resonance is positive, being the sum of
the positive left wings of the higher-order kinks, whereas the bias at the higher-order order resonances is negative, being dominated by the
negative right wing of the overarching 1:1 kink. As a result, the secular action of tides is to decelerate a planet rotating faster than $1\,n$,
and to accelerate a slower rotating planet (including a retrograde rotation). On the example shown in Fig \ref{kink.fig} (left) for the
5:2 resonance, we can see that
the negative bias can well be larger in absolute value than the amplitude of the corresponding kink, in which case the tidal torque is
negative (decelerating) everywhere around this resonance. However, this does not preclude the possibility of the planet to be captured
into this resonance. Fig. \ref{kink.fig} (right) depicts such a capture, simulated with these initial conditions: mean anomaly at time zero,
${\cal M}(0)=0$, initial rate of rotation, $\dot\theta(0)=2.52\,n$, sidereal angle of rotation, $\theta(0)=0$\footnote{All other notations used in this formula and throughout the paper are listed in Table \ref{nota.tab}.}. Thus, the capture of GJ 667Cc into a 5:2 resonance 
is possible with the set of estimated default
parameters (if not probable).

Following the ideas in \citep{mbe}, we herewith apply two different, independent methods to estimate the probabilities of capture into the
5:2 resonance. The first method is a brute-force integration of the ODE on a grid of initial conditions $\theta(0)$ for a fixed $\dot\theta(0)$
and a zero initial mean anomaly. The other method is to use a semi-analytical analogue of the capture probability formula derived by
\citep{gol68} for the simple constant- and linear-torque models. Using the first method, we performed 40 short-term integrations similar
to the one depicted in Fig. \ref{kink.fig} (right), with a fixed eccentricity $e=0.25$ and for a grid of initial rotation angles
$\theta(0)=\pi\,j/40$, $j=0,1,\ldots,39$. The time of integration in each case was 1000 yr, and the variable step of integration not larger than
$1.5\times 10^{-3}$ yr. Only 4 out of these 40 integrations resulted in capture, the others traversing this resonance. The estimated probability
of capture for the given set of parameters is $0.10\pm0.03$. The uncertainty of this and other numerically estimated probabilities are
simply the formal error on a Poisson-distributed sample estimate, given here only as a guidance. The uncertainty associated with the input
parameters may be more significant.

\begin{deluxetable}{lr}
\tablecaption{Explanation of notations \label{nota.tab}}
\tablewidth{0pt}
\tablehead{
\multicolumn{1}{c}{Notation}  &
\multicolumn{1}{c}{Description}\\
}
\startdata
$\xi$ & \dotfill moment of inertia coefficient \\
$R$ & \dotfill radius of planet \\
$T$ & \dotfill torque \\
$M_2$ & \dotfill mass of planet \\
$M_1$ & \dotfill mass of star \\
$a$ & \dotfill semimajor axis of planet \\
$r$ & \dotfill instantaneous distance of planet from star \\
$\nu$ & \dotfill true anomaly of planet \\
$e$ & \dotfill orbital eccentricity \\
$M$ & \dotfill mean anomaly of planet \\
$B$ & \dotfill second moment of inertia \\
$A$ & \dotfill third moment of inertia \\
$n$ & \dotfill mean motion, i.e. $2\pi/P_{\rm orb}$ \\
${\cal G}$ & \dotfill gravitational constant, $=66468$ m$^3$ kg$^{-1}$ yr$^{-2}$ \\
$\tau_M$ & \dotfill Maxwell time \\
$\mu$ & \dotfill unrelaxed rigidity modulus \\
$\alpha$ & \dotfill tidal lag responsivity \\
\enddata
\end{deluxetable}

\begin{figure}[htbp]
  \centering
  \plottwo{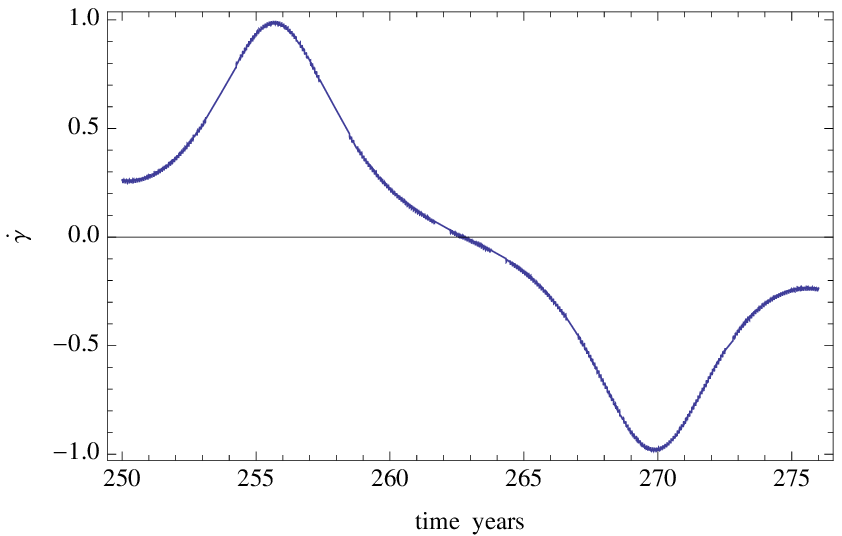}{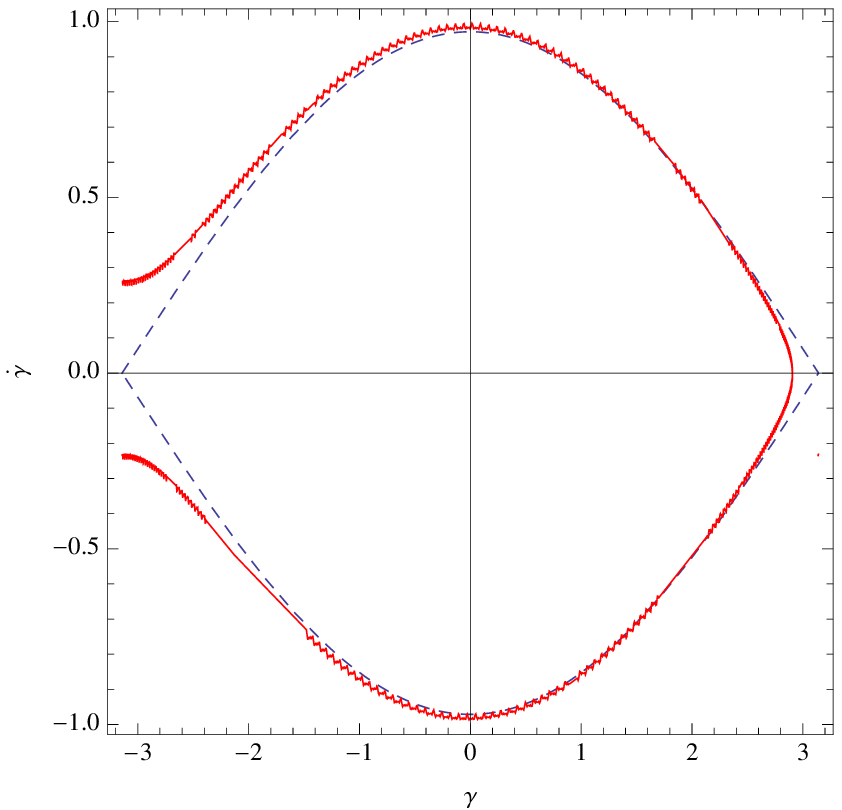}
\caption{Left: Simulated passage of GJ 667Cc's spin rate through the 5:2
spin-orbit resonance. Only two free libration cycles are shown, one immediately preceding the passage, and the other following it.
Right: The same two libration cycles as in the left panel, but as a separatrix trajectory in the $\{\gamma,\dot\gamma\}$ parameter plane.
The actual, accurately integrated trajectory is depicted with the bold line, and the assumed separatrix in the semi-analytical
calculation of capture probability with a dashed line. \label{libr.fig}}
\end{figure}

The other method is the semi-analytical calculation based on the energy balance consideration proposed by \citet{gold, gol68}.
An estimate of capture
probability is derived from the consideration of two librations around the point
of resonance $\dot\gamma=-\chi_{220q'}=2\dot\theta-(2+q')n=0$, i.e., the last libration with positive $\dot\gamma$ and the first
libration with negative $\dot\gamma$ for a slowing down planet. The angular parameter $\gamma=2\theta-(2+q'){\cal M}$ is introduced for
convenience. \citet{gold} assumed that the energy 
offset from zero
at the beginning of the last libration above the resonance is uniformly distributed between
$0$ and $\Delta E=\int\langle T\rangle\dot\gamma dt$. Then the probability of capture is
\eb
P_{\rm capt}=\frac{\delta E}{\Delta E},
\ee
with $\delta E$ being the total change of kinetic energy at the end of the libration below
the resonance. Thus, $\langle T\rangle\dot\gamma$ should be integrated over one cycle
of libration to obtain $\Delta E$, and over two librations symmetric around the resonance $\chi_{220q'}
=0$ to obtain $\delta E$. As a result, the odd part of the tidal torque at $q=q'$ doubles in
the integration for $\delta E$, whereas the bias vanishes; both these components are
involved in the computation of $\Delta E$. Denoting the bias and the kink components, respectively, $V$ and $W(\dot\gamma)$,
the capture probability is
\eb
P_{\rm capt}=\frac{2}{1+2\pi V/\int_{-\pi}^{\pi}W(\dot\gamma) d\gamma}
\label{prob.eq}
\ee
The integral
in this equation can be computed if we further assume that the trajectory in the vicinity of
resonance follows the singular separatrix solution of zero energy
\eb
\dot\gamma=2\,n\,\left[ \frac{3(B-A)}{C}G_{20q'}(e)\right]^\frac{1}{2}\cos\frac{\gamma}{2},
\label{separ.eq}
\ee
where $G_{lpq}(e)$ is the eccentricity function. The combination of Eqs. \ref{prob.eq} and \ref{separ.eq} makes for
a fast way of estimating the capture probability without the need of performing multiple integrations of differential equations.
It should also work for a rising spin rate, i.e., for an accelerating rotation.

Computation using this method with the default parameters of GJ 667Cc ($\tau_M=50$ yr, $e=0.25$) for the 5:2 resonance
yield a capture probability of 0.19. This estimate is significantly higher than the number (0.10) we obtained
by brute-force computations. The large discrepancy indicates that some of the assumptions used for one, or both,
of the methods is invalid or inaccurate. We investigated this problem in depth, and came to the following conclusion.
The weakest assumption in the semi-analytical method is the shape of the separatrix, Eq. \ref{separ.eq}. It assumes
that the libration curves begin and end at the resonant $\dot\gamma=0$. This may be a good approximation for a slow
tidal dissipation case, but it breaks for fast spin-downs, such as the one we are dealing with here. This conclusion is
illustrated by Fig. \ref{libr.fig}. The left panel of this Figure shows the variation of $\dot\gamma$ for $q'=3$ (i.e.,
5:2 resonance) while the planet is traversing the resonance, obtained by numerical integration. For a better
detail, only the two critical libration oscillations are displayed. We observe that the starting $\dot\gamma$ of the
pre-resonance libration is
significantly above the resonance, whereas the ending $\dot\gamma$ of the post-resonance libration is significantly
below it. These shifts are due to the substantial secular tidal torque acting on the planet, and the ensuing fast
deceleration. The shape of the separatrix trajectory is more clearly shown in the right part of Fig. \ref{libr.fig}
for the same pair of libration cycles. The pre-resonance libration is positive in this parametric plot, the
post-resonance libration is negative, and the planet moves clockwise along this trajectory. The actual, accurately
computed trajectory is shown with the solid red line, whereas the approximate trajectory described by Eq. \ref{separ.eq}
is shown as the dashed line. The actual trajectory does not make a closed loop, as the start and the end points
are separated by a gap. 

This discontinuity of the separatrix is always present, of course, due to the finite tidal dissipation of kinetic energy
during the two critical libration cycles. But in many cases, e.g., most of the bodies in the Solar system,
the tidal dissipation is so small that the departure from a closed, symmetric separatrix can be neglected. Furthermore,
for a constant or slowly varying with frequency tidal torque, this departure results in a small error, which can be
neglected. This is not the case for GJ 667Cc with our tidal model. As seen in Fig. \ref{kink.fig} (left), the kink
function is rapidly decreasing with tidal frequency on either side of the resonance. The "missing" part of the integrand
$W(\dot\gamma)\dot\gamma$ due to the gap may therefore bring about a significant change of estimated probability.
In order to test this idea, we extracted the appropriate segment of the $\dot\gamma$-curve from the integrated solution
and used this numerically quantified function in Eq. \ref{prob.eq} to compute the probability of capture. The resulting
probability is 0.14, which is much closer to the value $0.10\pm0.03$ estimated by brute-force integration. 

In the following analysis of probabilities of capture and of resonance end-states, we rely on the first, entirely numerical
way of estimation. For a given resonance $(2+q')$:2, $q'=0,1,\ldots$, and with a fixed $e$, we ran 40 simulations of
the spin rate, starting from a value of $\dot\theta$ above the resonant value, ${\cal M}(0)=0$, and 40 initial values of $\theta$
evenly distributed between 0 and $\pi$. If $N_c$ is the number of captures detected in a set of forty, the estimated
probability of capture is $N_c/40$. We further had to take into account that the eccentricity varies in a wide range
(Fig. \ref{ecc.fig}). When the planet's spin rate crosses a particular resonance, any phase value of the eccentricity
oscillation can be assumed equally probable. The probability of eccentricity to have a certain value at this time
can be approximated by dividing the full oscillation period into a number of intervals of equal length and computing the
median eccentricity for each interval. Table \ref{edistr.tab} gives the quantized probability distribution of $e$ estimated in
this fashion, using the results of numerical simulations described in \S \ref{cha.sec}. The top-range values are more likely
than the bottom-range values because the eccentricity curve is flat at the top. For each of the characteristic values of $e$,
we performed a set of 40 integrations of the spin-orbit differential equation on a regular grid of initial $\theta(0)$ and 
counted the number of captures. The estimated probability of capture for the given eccentricity was then one-fortieth of this
number. Each estimated probability of capture was multiplied by the corresponding probability of $e$, and the sum of these 10
numbers was the overall probability of capture for a random realization of $e$. This procedure was repeated for the 5:2, 2:1, and 3:2
resonances, resulting in a total of 1200 simulations.

Using this somewhat laborious method, we arrived at these probabilities of capture (on a single trial): 0.03 into 5:2, 0.23 into 2:1,
and 0.68 into 3:2. These numbers were obtained for the default $\tau_M=50$ yr and $(B-A)/C=5\times10^{-5}$. If we are more interested in the
current-state probabilities, these numbers need to be recomputed, taking into account that a planet locked into a higher resonance,
e.g., 5:2, can not ever reach a lower resonance, e.g., 2:1. The current-state probabilities are, obviously,
0.03 for 5:2, 0.22 for 2:1, and 0.51 for 3:2. The remaining trials, at a probability of 0.24, are certain to end up in the 1:1 resonance.

\begin{deluxetable}{lr|lr}
\tablecaption{Quantized probability distributions for planet GJ 667Cc and b. \label{edistr.tab}}
\tablewidth{0pt}
\tablehead{\multicolumn{2}{c}{planet c} & \multicolumn{2}{c}{planet b}\\
 \multicolumn{1}{c}{$e$}  & \multicolumn{1}{c}{$P_e$} & \multicolumn{1}{c}{$e$}  & \multicolumn{1}{c}{$P_e$}
}
\startdata
$0.061$ & 0.146 & 0.074 & 0.118 \\
$0.080$ & 0.072 & 0.095 & 0.060 \\
$0.100$ & 0.070 & 0.116 & 0.061 \\
$0.119$ & 0.070 & 0.138 & 0.061 \\
$0.138$ & 0.069 & 0.159 & 0.059 \\
$0.158$ & 0.071 & 0.180 & 0.069 \\
$0.177$ & 0.071 & 0.202 & 0.075 \\
$0.197$ & 0.088 & 0.223 & 0.093 \\
$0.216$ & 0.109 & 0.244 & 0.114 \\
$0.236$ & 0.232 & 0.266 & 0.289 \enddata
\end{deluxetable}

The probabilities of capture are known to depend on the degree of elongation $(B-A)/C$ and the Maxwell time $\tau_M$, 
which are quite uncertain. From our previous study on GJ 581d, we knew that the dependence on $(B-A)/C$ is much
weaker than on $\tau_M$. Bodies with smaller $(B-A)/C$, i.e., more spherical or axially symmetric, are more easily captured
into super-synchronous resonances. Generally, larger planets have smaller elongation parameters than the
smaller planets or moons. Our choice of this parameter is deemed conservative in terms of the capture probability
estimation. On the other hand, a warmer, less viscous planet with a smaller value of $\tau_M$ is much more
likely to be captured into super-synchronous equilibria. The average viscosity of the mantle is poorly known even
for the Solar system bodies, including the Moon. The amount of partial melt, in particular, may be crucially
significant for the spin-orbit evolution. The reverse is also true, in that the spin-orbit interactions define
the amount of tidal heat production, resulting, under favorable conditions, in a partial melt-down of the mantle,
or in a significant warming over the course of billions years. 

\section{Characteristic time of spin-down}
Assuming that in the distant past, the planet was rotating very fast in the prograde sense, how long does it take to
spin down and fall into one of the spin-orbit resonances? This can be assessed through a parameter called
the characteristic spin-down time, customarily defined as
\eb
\tau_{\rm spin-down}=\frac{\dot\theta}{|\ddot\theta|},
\label{tau.eq}
\ee
where $\ddot\theta$ is the angular acceleration caused by the secular component of the tidal torque. This time
parameter should not be confused with the actual time for the planet to decelerate from a certain initial spin rate
and fall into a resonance, which is normally shorter. Indeed, Eq. \ref{tau.eq} allows us to quickly compute the
{\it instantaneous} angular acceleration $\ddot\theta$ and the corresponding spin-down time for a given spin rate $\dot\theta$ using
known analytical equations for the secular tidal torque. But the angular acceleration by itself is a nonlinear function
of $\dot\theta$; therefore, the rate of spin-down grows faster as the planet decelerates. 

\begin{figure}[htbp]
  \centering
  \plotone{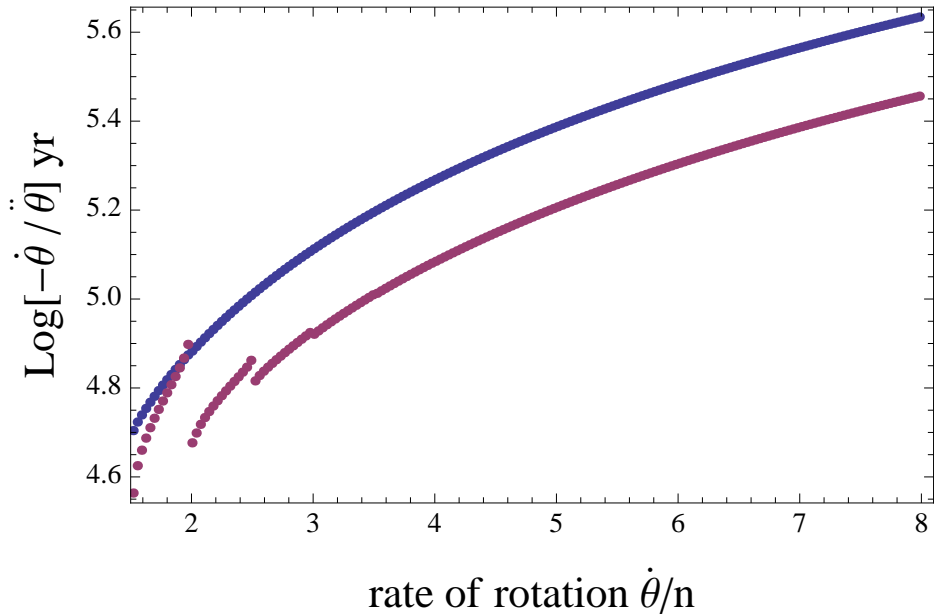}
\caption{Characteristic times of tidal spin-down of the planet GJ 667Cc for two values of orbital eccentricity:
0.06 (upper curve) and 0.24 (lower curve).
\label{spindown.fig}}
\end{figure}

The instantaneous characteristic spin-down times as functions of the
spin rate are shown in Fig. \ref{spindown.fig} for two values of $e$, which bracket the range of its
variation, the upper curve corresponding to $e=0.06$ and
the lower curve to $e=0.24$. In both cases, we assumed the present-day observed values of semimajor axis
(Table \ref{para.tab}). We find that for $\dot\theta < 8\,n$, the spin-down times are well within 1 Myr, which is
much shorter than the presumed life time of the planet. The sign of $\ddot\theta$ changes to positive for spin rates slower
than the mean orbital motion, $\dot\theta < 1\,n$, i.e., the planet spins up with such slow rotation rates, including retrograde rotation.
The apparent discontinuities of the lower curve correspond to main supersynchronous resonances, which the planet either traverses
quite quickly or becomes entrapped in. The short spin-down times suggest that the planet was captured into the current resonant
state as long as a few Gyr ago. This may also indicate that the semimajor axes and the separation between the planets were
different when this capture happened, because the energy for tidal dissipation is drawn from the orbital motion when the spin rate
is locked. The orbital evolution of two-planet systems with significant tidal dissipation locked both in a MMR and a spin-orbit resonance
is a complex problem, which lies beyond the scope of this paper.

\section{Likely spin-orbit state of GJ 667Cb}
\label{b.sec}
According to the data in Table \ref{para.tab}, the planet b is much closer to the host star and is somewhat more massive than the planet c.
As the polar component of the tidal torque scales as $T_z\propto R^5/r^6$ \citep{ema}, where $R$ is the radius of the planet and r is the distance to
the host star, the tidal forces on planet b should be at least 200 times stronger than on planet c. This, however, does not necessarily imply
higher probabilities of capture into super-synchronous spin-orbit resonances. To understand why this is the case, the analogy of the capture
process to a rotating driven pendulum with damping may be useful \citet{gol68, makti}. The overall (negative) bias of the tidal torque
acts as a weak driving force against the initial prograde rotation of the pendulum, whereas the frequency-dependent, and highly nonlinear in our
case, component of the torque acts as friction. The pendulum gradually slows down in its rotation and, inevitably, it is no longer able to come
over the top. On the first backward swing, the bias will assist it in passing the top in the opposite direction, while the friction will
further diminish the amplitude of oscillation. The subtle balance between these components become crucial in whether the pendulum can traverse
the top point and commence rotating in the retrograde sense, or it becomes locked in the gradually diminishing swings around the point of 
stable equilibrium. The probabilistic nature of these outcomes originates from the finite range of possible positions when it is stalled 
in the vicinity of the top point. A much stronger tidal force increase the bias and the frequency-dependent (friction) components in the
same proportion, but the balance between them is mostly affected by the orbital eccentricity. Since the eccentricity of the two planets
are not too different on average (Fig. \ref{ecc.fig}), the capture probabilities may be close too.

Here we compute the planet b capture probabilities for the resonances 3:2, 2:1 and 5:2 essentially repeating the steps described
in \S\ref{c.sec}. We assume the same Maxwell time, $\tau_M=50$ yr, as for planet c. On a grid of regularly spaced points in eccentricity
between 0.074 and 0.266, 40 simulations with uniformly distributed initial libration angles are performed, starting with a spin rate
above the resonant value. The number of captures is counted in each batch of 40 simulations, with the total number of batches being
300. The estimated probabilities are weighted with the binned probabilities of eccentricity given in Table \ref{edistr.tab} and
summed up for each resonance. The probabilities of individual capture events are: 0.82 for 3:2, 0.32 for 2:1, and 0.10 for 5:2
resonances. Capture into the 1:1 resonance is certain. If the planet evolved from high prograde spin rates, which seems to be the
likeliest scenario judging from the Solar planets, it can reach a lower resonant state only if it traversed all the higher
resonances. Taking into account the compounding conditional probability, the probabilities of the end-states are: 0.10 for 1:1,
0.51 for 3:2, 0.29 for 2:1, and 0.10 for 5:2 resonances.

Again, as in the case of planet c, the most likely state for planet b is the 3:2 spin-orbit resonance. The distribution
of probabilities for b is shifted toward higher orders of resonance because of the slightly larger average eccentricity.
If the planet originally had a retrograde spin (such as Venus in the Solar System), the only long-term stable state is
the complete synchronization of rotation.

\section{The rate of tidal heating}
An exoplanet of terrestrial composition captured into a spin-orbit resonance continues to dissipate the orbital kinetic
energy through the tidal friction inside its body. If the resonance is not synchronous, the spin rate differs from the orbital
rate, resulting in a constant drift of the tidally-raised bulge across the surface of the planet. The internal shifts of the
material cause the mantle to warm up. Naively, one would expect that faster motions of the tidal bulge should bring about
higher rates of dissipation and, therefore, more vigorous production of tidal heat. This indeed follows from the commonly
used Constant Time Lag (CTL) model of tides \citep[e.g.,][]{leco}, which is also commonly misapplied to rocky planets.
There are two crucial defects of such models as applied to rocky exoplanets: 1) the actual rheology of earth-like solids unambiguously
implies a declining with frequency {\it kvalitet}\footnote{Kvalitet stands for "quality" in Danish, which we use here as a more general term for the
customary tidal quality factor $Q$ in the literature.} at high perturbation frequencies; 2) self-gravity strongly limits the amplitudes
of tides on large exoplanets and stars. On the other hand, the tidal bulge is stationary with respect to the planet's body if
the planet is locked into a synchronous rotation {\it and} both the orbital eccentricity and obliquity of the equator are exactly zero;
in such a hypothetical case, the tide is on, but there is no tidal dissipation of energy or heat production. 

We have determined that the most likely state of GJ 667Cc is a 3:2 resonance. In this state, the planet makes a full turn around its axis
3 times for every 2 orbital periods with respect to distant stars, but only one turn for every two orbital periods with respect to the host star.
In other words, the day on this planet is likely to be $28.1\times 2=56.2$ d long. The main semi-diurnal tidal mode will have a
period of 28.1 d. Apart from the average prograde motion of the tidal bulge, relatively small oscillations of the tidal perturbation
should be expected from the longitudinal and latitudinal librations. Physically, the librations can be separated into two categories,
the forced librations caused by a variable perturbing force, and the free librations caused by an initial excess of kinetic energy.
The latter kind of librations is expected to damp relatively quickly for Mercury-like planets \citep{peal}, with a characteristic damping time
orders of magnitude shorter than the planet's life time. The remaining forced librations in a two-body system are solely due to the
periodical orbital acceleration of the perturber on an eccentric orbit, and their amplitude is quite small for massive super-earths.
Additional harmonics of forced libration should be expected for GJ 667C planets due to their mutual interaction. It turns
out that the periodical terms of the tidal torque are completely insignificant in comparison with the secular modes in this case.

\begin{figure}[htbp]
  \centering
  \includegraphics[angle=0,width=0.95\textwidth]{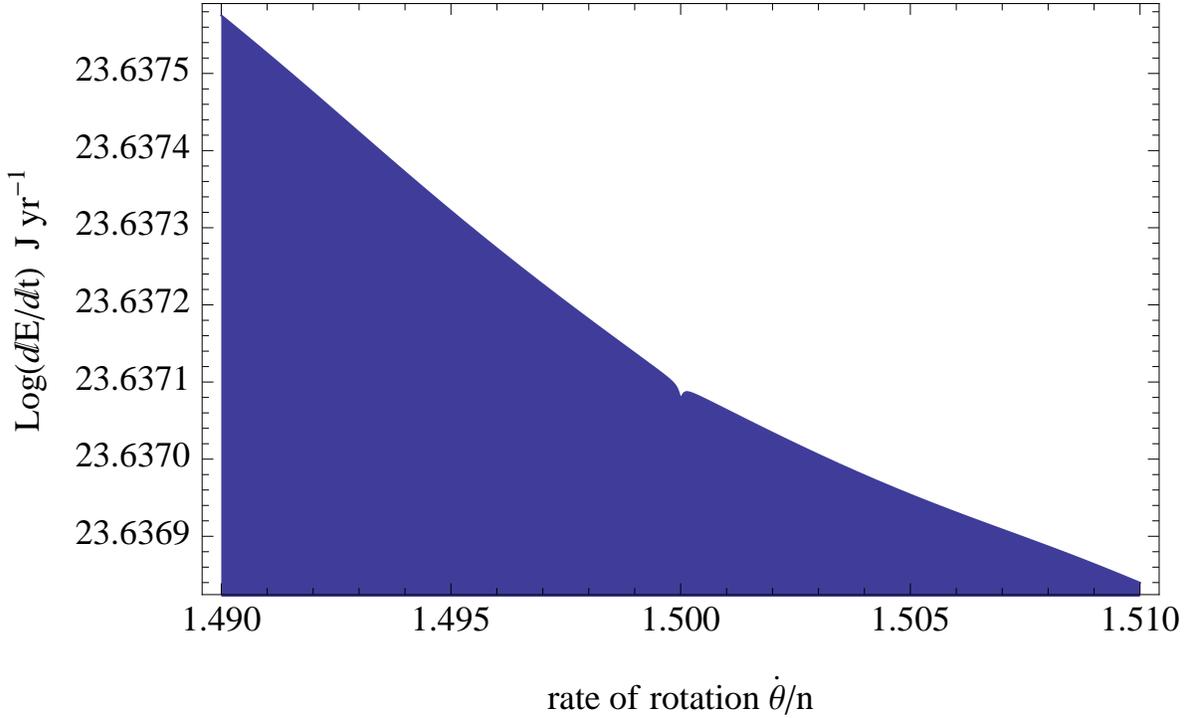}
\caption{The rate of tidal energy dissipation inside GJ 667Cc in the vicinity of the 3:2 spin-orbit
resonance, which we find the most likely state for this planet.   \label{heat.fig}}
\end{figure}

The formula to compute the rate of energy dissipation is taken from \citet[][Eq. 2]{mamoon}, which is an adaptation of the
seminal work by \citet[][in particular, Eq. 31]{peca} for the synchronously rotating Moon. The latter equation was
essentially obtained following the geometrical consideration by \citet{kau}. This formalism can be used for any eccentricity
and any spin rate, as long as the secular components of the tidal dissipation are concerned. It should be noted that the $lmpq$ terms in the 
series given by \citet{peca} appear to come with different signs. In practice, their absolute values should be used instead,
because each tidal mode can only increase the dissipation, independent of the sign of the corresponding tidal torque.
Alternatively, one can consider the factor $Q_{lmpq}$ to be an odd function of tidal frequency.
With these important corrections, the resulting dissipation rate is significantly higher than what would have been obtained
with the approximate equation truncated to $O(e^2)$. Fig. \ref{heat.fig} shows the rate of energy dissipation computed
as a function of spin rate in the vicinity of the 3:2 resonance, for the default parameters of GJ 667Cc and $e=0.15$.
The gently sloping curve is only marked with a tiny dent at the resonant spin rate. Therefore, the exact shape and amplitude
of longitudinal librations is not important for the estimated rate of dissipation. The dependence on eccentricity is not strong either,
ranging from $10^{23.53}$ J yr$^{-1}$ for $e=0.05$ to $10^{23.77}$ J  yr$^{-1}$ for $e=0.22$. For the median
value of eccentricity $e=0.18$, the estimated energy dissipation rate is $10^{23.7}$ J yr$^{-1}$.

We are adopting a value of 1200 J kg$^{-1}$ K$^{-1}$ for the planet's heat capacity from \citep{beho}. The average
rise of temperature for the entire planet is $1.6\cdot 10^{-5}$ K yr$^{-1}$. At this rate, the mantle should reach
the melting point of silicates in less than $0.1$ Gyr. More accurate calculations than we are able to carry out
in this paper would require a careful modeling of the mantle convection effects, radiogenic heating, and heat transfer.
But given the warm-up time that is much shorter than the life time of M dwarfs, it will not be too bold to say that
a considerable degree of melting and structural stratification should have occurred on planet GJ 667Cc. In that respect,
the situation is reminiscent of Mercury in the Solar System, which has a massive molten core extending up to
0.8 of its radius. The spin-down time for Mercury is of the order of 10$^7$ yr \citet{noel}, suggesting that the planet
has been in the current 3:2 spin-orbit resonance for billions of years. The molten core therefore formed after the
capture into this resonance. The tidal dissipation rate at a super-synchronous rotation resonance becomes of
an overarching importance for our understanding of the structure and destiny of inner planets subject to relatively
strong tidal forces.

If the tidal dissipation rate suggests that at least a partial melt should have occurred on the planet GJ 667Cc,
similar calculations for the planet GJ 667Cb leave little doubt that the planet should be completely molten.
For a median eccentricity $e=0.18$ and the same heat capacity, the estimated heat production is 1.1 K in just 100 yr.
The planet should quickly become a ball of molten magma. The most likely scenario for such close-in planets
trapped in resonances seems to be overheating and destruction. But something obviously has prevented this planet from complete
evaporation. This problem requires a dedicated and accurate study; here, we only offer some possibilities that could change
the conclusions. When the temperature of a rocky planet rises, its rheology changes too. In particular, the Maxwell
time in the expression for kvalitet is quite sensitive to the average temperature. 
The probabilities of capture into spin-orbit resonances estimated in Sections \ref{c.sec} and \ref{b.sec} can only
become higher for a warmer planet due to the shortening of its Maxwell time. If the planets were already heated up
by the time their spin rates reached the lower commensurabilities with the orbital motions, they are even more likely
to have been captured and to have remained in the higher spin-orbit resonances.
For partially molten or partially liquid
bodies, the Maxwell time may be comparable to the period of rotation, or shorter, which is probably the case for
Titan (F. Nimmo, priv. comm.). This may drastically change the frequency-dependent terms of tidal dissipation. At the extreme,
a ball of water or a planet with a massive ocean is likely to have a principally different tidal response than solid
bodies \citep{tyl}. Besides, a liquified planet can lose its ``permanent" figure and become nearly perfectly spherical or
oblate, radically changing the conditions of continuous entrapment in the resonance. A combination of these events, speculatively,
can create a seesaw effect, when an eccentric solid planet is liquified by the tidal dissipation, which causes the
rate of dissipation to drop by orders of magnitude, followed by a slow cooling, solidification of the surface layers,
acquiring a permanent figure shaped by the tidal interaction, bringing up an episode of strong tidal heating and
melt-down, and so on, ad infinitum.

\section{Conclusions}
The rapidly growing class of detected super-Earths, which may reside in the habitable zone and, thus, harbor
life in the biological forms familiar to us, sets new objectives and motivations for the interpretation of
planetary spin-orbital dynamics and the theory of tides. The history of planet's rotation and the tidal
dissipation of kinetic energy in the long past is certain to play a crucial role in the formation (or the absence of such)
of liquid oceans and gaseous atmospheres. This study is ridden with uncertainties of both observational and
theoretical kind. Earthlings are blessed with a rapid and stable rotation of their home planet, but are
the potentially habitable super-Earths similarly hospitable in this respect? Looking at the better known
Solar planets and satellites, which objects are the closest analogy to the massive super-Earths orbiting
near their M-type hosts?

Judging from the observational data we have today, and making use of the much improved tidal model for rocky planets,
the super-Earths seem to be more similar to the tiny Mercury than to the Earth, as far as their spin-orbit dynamics
is concerned. Mercury, making exactly 3 sidereal rotations per one orbital period \citep{pett}, is the only planet in
the Solar system captured into a supersychronous resonance. This resonance happens to be the most likely outcome
of Mercury's spin-orbit evolution even without the assistance of a liquid core friction, due to the relative proximity
of the planet to the Sun and its considerable eccentricity, which could have reached even higher values in the past
\citep{corla04}. Mercury also has rather short characteristic times of spin-down of the order of $10^7$ yr
\citep{noel}, which indicate that the massive molten core was formed after the capture event. The same circumstances
define a supersynchronous resonant rotation as the most probable state of both GJ 667C planets. 

The characteristic spin-down time for the planet GJ 667Cc is of the order of 1 Myr, which is even shorter than that
for Mercury. The planet is certain to be in one of the low-order resonances, that is, its current rotation with respect
to the host star is very slow, if any. Due to the massiveness of this exoplanet, the longitudinal forced librations
are likely insignificant for the considerations of its habitability. However, being locked into a mean-motion resonance
with planet b, the orbit undergoes rapid, high amplitude variations having, undoubtedly, a significant impact on the
circulation of the hypothetical atmosphere and climate. The slow relative rotation and a long solar day generally imply rather harsh
conditions on the surface; at the same time, a combination of the significant eccentricity, obliquity of the equator (which is
ignored in this paper), and longitude of the vernal equinox may provide for well-shielded areas on the surface with
favorably stable and moderate insolation \citep{dobr}.

Perhaps even more threatening in terms of habitability
is the high rate of tidal heating that we estimate for the planet GJ 667Cc. The estimated rise
of average temperature by 1.6 K per $10^5$ yr is likely to cause a partial or complete melting of the planet's mantle.
It remains to be investigated if there are any safety mechanisms in the physics of tidal dissipation that can
automatically prevent a super-Earth on an eccentric orbit from overheating. The rate of tidal dissipation in resonantly spinning super-Earths
happens to be weakly dependent on orbital eccentricity, which is not very accurately determined by the RV planet detection technique. 
For example, the rate of dissipation at $e=0.05$ is smaller by just $dex(0.1)$ compared to that at $e=0.15$. Likewise, the
unknown parameters $\tau_M$ and $(B-A)/C$ have a very limited impact on the accuracy of this estimation. The radius of the planet
and the average distance to the host star appear to be the defining parameters, due to the explicit proportionality to
$R^5/a^6$. The estimated tidal heating of the inner planet GJ 667Cb is yet higher by roughly three orders of magnitude,
leaving us in wonder how such close-in, strongly interacting bodies can avoid seemingly inescapable death by fire.

This research has made use of the Washington Double Star Catalog maintained at the U.S. Naval Observatory.

\end{document}